\documentclass[11pt]{article}
\usepackage{times}
\usepackage{sidecap}
\usepackage{geometry}
\usepackage{floatrow}
\usepackage{float}
\usepackage{hyperref}
\usepackage{graphics,graphicx}
\usepackage{subcaption}
\usepackage{epsf,wrapfig}

\usepackage{natbib,bm}

\usepackage{paralist}
\let\olditem\item
\renewenvironment{thebibliography}[1]{%
  \section*{\refname}
  \let\par\relax
  \renewcommand{\item}[1][]{\olditem[\textbullet]}%
  \inparaenum}{\endinparaenum}

\newcounter{aff}

\DeclareGraphicsExtensions{.jpg,.pdf,.png,.eps}

\usepackage{caption}
\usepackage{xcolor}

\usepackage{tcolorbox}

\bibpunct{(}{)}{;}{a}{}{,}

\usepackage{soul}

\definecolor{lightcyan}{rgb}{0.7, 0.8, 1.0}
\definecolor{darkcyan}{rgb}{0.0, 0.0, 0.7}

\definecolor{boxcolor}{RGB}{193, 229, 255}
\definecolor{figboxcolor}{RGB}{193, 229, 255}
\definecolor{boxcolorgray}{RGB}{227, 227, 227}
\definecolor{darkcyan}{rgb}{0.0, 0.0, 0.7}

\sethlcolor{boxcolorgray}

\def\gtrsim{\lower 2pt \hbox{$\, \buildrel {\scriptstyle >}\over
{\scriptstyle \sim}\,$}}
\def\lesssim{\lower 2pt \hbox{$\, \buildrel {\scriptstyle <}\over
{\scriptstyle \sim}\,$}}

\def\approxlt{\lower.2em\hbox{$\buildrel < \over \sim$}}
\def\approxgt{\lower.2em\hbox{$\buildrel > \over \sim$}}

\def\be{\begin{equation}}
\def\ee{\end{equation}}
\def\bea{\begin{eqnarray*}}
\def\eea{\end{eqnarray*}}

\def\ci{[C{\sc i}]}
\def\cii{[C{\sc ii}]}

\def\oii{[O{\sc ii}]}

\def\jref@jnl#1{{\rm#1}}
\def\aj{\jref@jnl{AJ}}			
\def\araa{\jref@jnl{ARA\&A}}		
\def\apj{\jref@jnl{ApJ}}			
\def\apjl{\jref@jnl{ApJ}}		
\def\apjs{\jref@jnl{ApJS}}		
\def\ao{\jref@jnl{Appl.~Opt.}}		
\def\apss{\jref@jnl{Ap\&SS}}		
\def\aap{\jref@jnl{A\&A}}		
\def\aapr{\jref@jnl{A\&A~Rev.}}		
\def\aaps{\jref@jnl{A\&AS}}		
\def\azh{\jref@jnl{AZh}}			
\def\baas{\jref@jnl{BAAS}}		
\def\jrasc{\jref@jnl{JRASC}}		
\def\memras{\jref@jnl{MmRAS}}		
\def\mnras{\jref@jnl{MNRAS}}		
\def\pra{\jref@jnl{Phys.~Rev.~A}}	
\def\prb{\jref@jnl{Phys.~Rev.~B}}	
\def\prc{\jref@jnl{Phys.~Rev.~C}}	
\def\prd{\jref@jnl{Phys.~Rev.~D}}	
\def\pre{\jref@jnl{Phys.~Rev.~E}}	
\def\prl{\jref@jnl{Phys.~Rev.~Lett.}}	
\def\pasp{\jref@jnl{PASP}}		
\def\pasj{\jref@jnl{PASJ}}		
\def\qjras{\jref@jnl{QJRAS}}		
\def\skytel{\jref@jnl{S\&T}}		
\def\solphys{\jref@jnl{Sol.~Phys.}}	
\def\sovast{\jref@jnl{Soviet~Ast.}}	
\def\ssr{\jref@jnl{Space~Sci.~Rev.}}	
\def\zap{\jref@jnl{ZAp}}			
\def\nat{\jref@jnl{Nature}}		
\def\iaucirc{\jref@jnl{IAU~Circ.}}
\def\aplett{\jref@jnl{Astrophys.~Lett.}}
\def\apspr{\jref@jnl{Astrophys.~Space~Phys.~Res.}}
\def\bain{\jref@jnl{Bull.~Astron.~Inst.~Netherlands}}
\def\fcp{\jref@jnl{Fund.~Cosmic~Phys.}}
\def\gca{\jref@jnl{Geochim.~Cosmochim.~Acta}}
\def\grl{\jref@jnl{Geophys.~Res.~Lett.}}
\def\jcp{\jref@jnl{J.~Chem.~Phys.}}	
\def\jgr{\jref@jnl{J.~Geophys.~Res.}}	
\def\jqsrt{\jref@jnl{J.~Quant.~Spec.~Radiat.~Transf.}}
\def\memsai{\jref@jnl{Mem.~Soc.~Astron.~Italiana}}
\def\nphysa{\jref@jnl{Nucl.~Phys.~A}}
\def\physrep{\jref@jnl{Phys.~Rep.}}
\def\physscr{\jref@jnl{Phys.~Scr}}
\def\planss{\jref@jnl{Planet.~Space~Sci.}}
\def\procspie{\jref@jnl{Proc.~SPIE}}

\def\ion#1#2{#1$\;${\small\rm\@Roman{#2}}\relax}

\newcount\lecurrentfam
\def\LaTeX{\lecurrentfam=\the\fam \leavevmode L\raise.42ex
\hbox{$\fam\lecurrentfam\scriptstyle\kern-.3em A$}\kern-.15em\TeX}
\def\sizrpt{
(\fontname\the\font): em=\the\fontdimen6\font, ex=\the\fontdimen5\font
\typeout{
(\fontname\the\font): em=\the\fontdimen6\font, ex=\the\fontdimen5\font
}}
\definecolor{dark}{rgb}{ 0.34, 0.09, 0.49}

\usepackage{fancyhdr,url}
\usepackage{transparent}
\usepackage{tablefootnote}
\usepackage{mathpazo}

\begin{document}

\noindent
\begin{raggedright}
{\Huge\bfseries The Magnetized Universe: \\[0.5em]
\Large  From planet-forming disks to the high-redshift universe
}\end{raggedright}\\

\noindent

Josep M Girart$^{1,*}$, Gary Fuller$^2$, Wouter Vlemmings$^3$, Jianhang Chen$^4$, 
Tom Bakx$^3$, Maria T Beltr\'an$^5$, Henrik Beuther$^6$, Leindert Boogaard$^7$,
Tao-Chung Ching$^{8}$,
Paulo Cortes$^{9}$,
Maryvonne Gerin$^{10}$,
Bo Huang$^{11}$,
Woojin Kwon$^{12}$, 
Shih-Ping Lai$^{13}$, 
Valentin Le Gouellec$^{1}$,
Hauyu Baobab Liu$^{14}$, 
Junhao Liu$^{15}$,
Enrique Lopez-Rodriguez$^{16}$,
Frédérique Motte$^{17}$,
Rosita Paladino$^{18, 21}$, 
Aina Palau$^{19}$,
Kate Pattle$^{20}$, 
Gerg\H{o} Popping$^{21}$,
Keping Qiu$^{22}$,
Ramprasad Rao$^{23}$, 
Víctor Revilla$^{24}$,
Sarah Sadavoy$^{25}$,
Raghvendra Sahai$^{26}$
Álvaro Sánchez-Monge $^1$, 
Patricio Sanhueza$^{27}$, 
Asako Sato$^1$,
Kei Tanaka$^{28}$,
Maxime Tanious$^{1}$,
Fengwei Xu$^{^6}$,
Qizhou Zhang$^{22}$

\vspace{1em}
{
\noindent $^*$ girart@ieec.cat \\
\noindent Affiliations are listed at the end of the text
}

\begin{tcolorbox}[boxsep=3pt,left=0pt,right=0pt,top=1pt,bottom=1pt,colframe=boxcolor,colback=boxcolor]
\justify
\textbf{Abstract:} 
The interstellar medium (ISM) is magnetized at all scales and phases in our Galaxy and other galaxies across the Universe. 
Spectro-polarimetric millimetre observations are essential to probe the role of the interstellar magnetic fields in the dynamics of the cold universe, and how this affects the star formation process, and planet formation in nearby star forming regions. 
Such observations are also essential for understanding magnetic fields in active stars and the shaping and evolution of the circumstellar environments of evolved stars.
While ALMA has been revolutionary in advancing our knowledge about magnetic fields in the ISM, 
the limitations in calibration, especially for circular polarization, and sensitivity for large samples, have hindered robust measurements of field strengths, which are essential for fully assessing their role in ISM dynamics.
An ALMA2040 that broadens spectro-polarimetric capabilities to include robust circular calibration and increases sensitivity with more collecting area will enable crucial observations of magnetic fields across the Universe and over a wide range of spatial scales to fully ascertain the influences of magnetic fields from galaxy assembly to forming planets.
\end{tcolorbox}

\vspace{0.2em}
\vfill

\begin{tcolorbox}[boxsep=2pt,left=0pt,right=0pt,top=7pt,bottom=0pt,colframe=boxcolorgray,colback=boxcolorgray]
\transparent{0.9}
\begin{center} 
\includegraphics[width=1.0\textwidth]{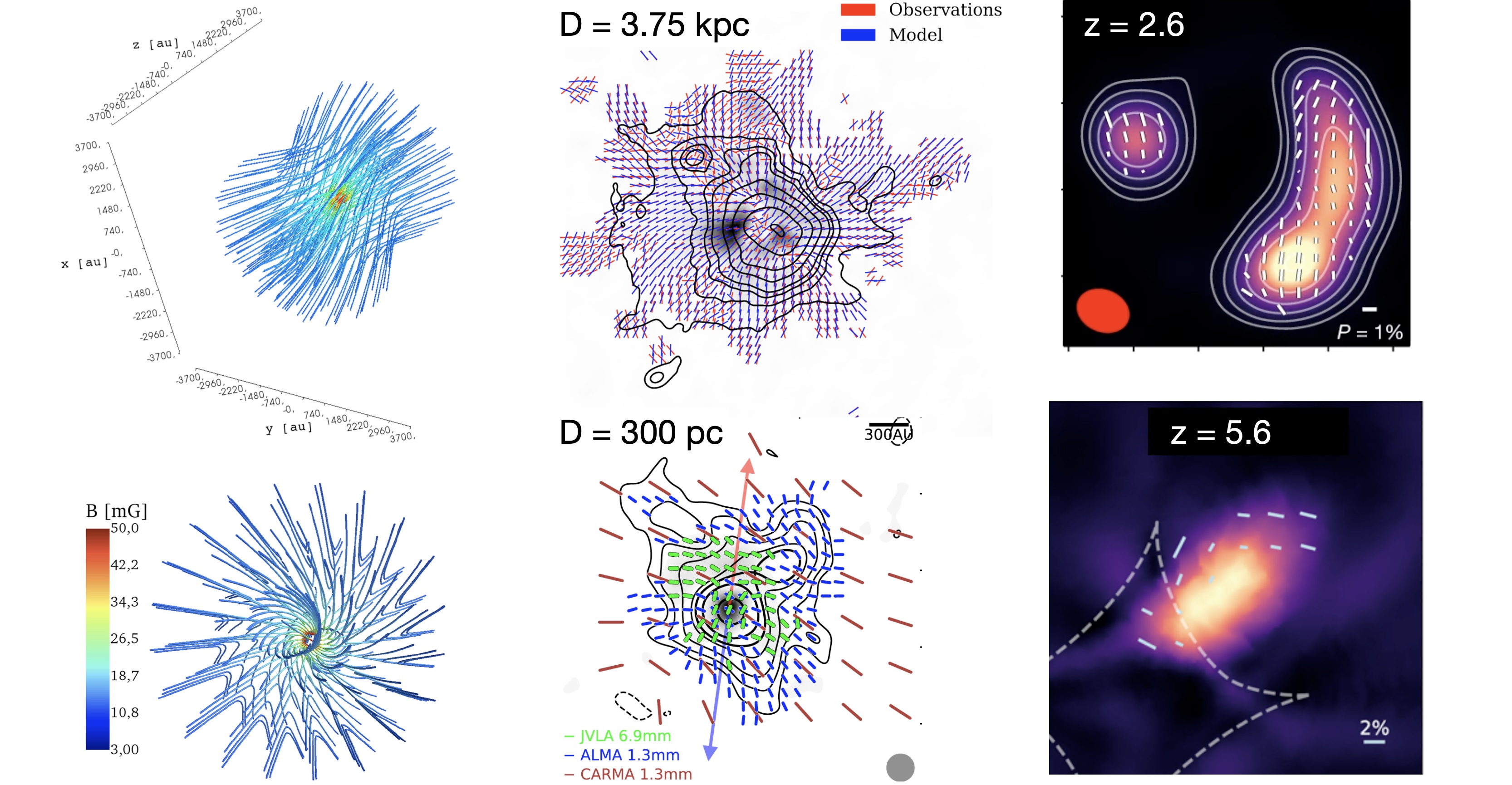} 
\end{center}
\justify
Figure 1: Composite from \cite{Beltran2019, Ko2020, Geach2023Natur, Chen2024}, . \textit{Left: } 3-D MHD simulations of the magnetic fields around G31.41+0.31, a massive star forming region. 
\textit{Middle: } Millimetre dust polarization observations of a massive star forming clump (top) and a protostar (bottom). \textit{Right:} ALMA images of dust polarization in high-redshift galaxies. 
\end{tcolorbox}


\section{Scientific context and open questions}

Magnetic fields  (B-fields) thread the entire interstellar medium (ISM) of our Galaxy and other galaxies across the Universe (see Fig.~1) and play an important role in many stages of stellar evolution. Characterizing magnetic-field strength and structure across the different phases of the ISM and across stellar evolution is essential for understanding their role in interstellar and (circum)stellar processes. At millimetre wavelengths, the main observational signatures used to study  magnetic fields are spectral line polarization and dust polarization. Molecular lines (and atomic lines such as \ci\ at 492 GHz) are sensitive to the magnetic field, splitting their transitions into magnetic sublevels, producing linear and circular polarization. For molecular radicals such as CN, CCS or CCH, the splitting is strong enough that circular polarization can arise at levels of $\sim1$\%. Detection of this Zeeman effect is the only way to directly quantify the B-field strength, since the net circular polarization is proportional to the strength along the line-of-sight \citep[see][]{Crutcher2019}. 
Thus, Zeeman observations are of utmost importance to directly test the relevance of magnetic fields in molecular clouds, star formation processes, stellar evolution and disk formation and evolution. 
For the vast majority of molecular species, 
the Zeeman effect is not observable, but under some conditions linear polarization can arise through the Goldreich-Kylafis (G-K) mechanism \citep[e.g., anisotropic radiation or moderate optical depth lines][]{GK1982}. This mechanism traces the B-field component on the plane-of-the-sky, producing linear polarization either parallel or perpendicular to the magnetic field direction, depending on the physical conditions of the emitting region \citep[e.g.][]{Girart1999, Lai2003}. 
Fast spinning elongated interstellar grains that align perpendicular to the local magnetic field produce linear polarization perpendicular to the magnetic field direction \citep{Lazarian2007}.  Note that neither the G-K effect nor dust polarization provide a direct measurement of the field strength.  This is inferred usually through indirect statistical approaches, such as the Davis-Chandrasekhar-Fermi method \citep[e.g.][]{DCF1953, Liu2022} or by modelling directly the expected B-field configuration \citep[e.g.][]{Frau2011}. 
In addition, there are other effects that one needs to take into account when studying the polarized signal for both continuum and spectral line observations, such as dust self-scattering \citep[produced by large grains][]{Kataoka2015, Yang2017}, dust alignment by mechanically/aerodynamically processes \citep{Lin2024b}, anisotropic resonant scattering effect in spectral lines \citep[conversion from linear to circular polarization;][]{Houde2013}, and linear polarization in molecular ions generated by ambipolar drifts \citep{Lankhaar2020}.

Currently, facilities like ALMA allow linear and circular polarization observations for some bands and under some restrictions (see Section~\ref{TechReq}). Nonetheless, the high sensitivity and high image fidelity have allowed a significant advance in the study of magnetic fields using the ALMA polarization capabilities, mostly from dust polarization observations. The results obtained from molecular line polarization are more scarce, and no Zeeman detection has been reported yet. 
\begin{itemize}
\item \underline{Planet Formation: }
Magnetic fields play a fundamental role in governing both the initial formation and long-term evolution of accretion disks around young stellar objects \citep[e.g.,][]{Maury22}. However, it is not clear whether they play a significant role in planet formation. Measuring magnetic fields in disks with ALMA is challenging because of the expected complex density, chemical and magnetic field properties \citep[e.g.,][]{Mazzei2020}, as well of the dominance of dust self-scattering \citep{Kataoka2017, Bacciotti2018, Lin2024, Liu2024, Yang2025}.  ALMA has not yet been able to detect magnetic fields in disks via Zeeman or G-K polarization observations \citep{Vlemmings2019, Harrison21, Stephens2020}, although there may be an indirect way to measure them \citep{Teague2025}. Recently, it has been shown that dust segregation within a disk allows dust polarization to partially trace the magnetic field \citep{Ohashi2025}.
\item \underline{Magnetized Star Formation in the Milky Way:} A wealth of publications has resulted from linearly polarized dust emission. 
On one side, deep observations on specific targets characterize B-fields with high fidelity, allowing the quantitative comparison of the observations with the theoretical models 
\citep[e.g.,][]{Maury2018, Kwon2019, Beltran2019, Beltran2024, Saha2024}. 
On the other side, several programs have targeted a statistically significant number of young stars in order to characterize the general trends across high- and low-mass star-forming regions, for example: BOPS -- B-field Orion Protostellar Survey \citep{Huang2024, Huang2026SciA}, MagMaR -- Magnetic fields in Massive star-forming Regions \citep{Sanhueza2021, Sanhueza2025} and ALPPS -- ALMA Perseus Polarization Survey \citep{Cortes2025}. On ALMA scales, gravity dominates over magnetic fields, with turbulence in massive cores also playing an important role \citep{Liu2026NatAs},  but they appear to play a key role funnelling the infalling gas through magnetically structured accretion streamers \citep{Huang2026SciA}.

Less explored with ALMA has been the detection of the G-K effect in molecular tracers such as CO.  Studies with the G-K effect have been limited by current ALMA capabilities, but these early detections suggest that with a substantial sensitivity boost, one could place meaningful constraints on the 3D magnetic field geometry \citep{Cortes2021, Ching2026NatCom}.

\item \underline{Stellar evolution:}
During the later stages of evolution, linear polarization due to the G-K effect has been  mapped in various molecular lines for a handful of nearby (post-)AGB stars as well as the supergiant VY CMa \citep{Vlemmings2017, Vlemmings2024}. These observations reveal that large-scale magnetic fields thread the circumstellar envelope. In the near-circumstellar environment, ALMA observations have revealed an energetic ejection, similar to a coronal mass ejection, from the AGB star Mira A, providing strong evidence that magnetic fields contribute to evolved star mass loss \citep{Vlemmings2026}. For main-sequence stars, direct ALMA polarimetry is only now emerging with the 2025 Band-2 science-verification observations of the magnetic B2.5V star HR 5907 demonstrating ALMA's capability to measure its strong, variable circular polarization.

However,  because of intrinsic calibration uncertainties, ALMA observations so far have been unsuccessful in determining the polarization of ultracool brown dwarfs \citep{Hughes2021} to determine the emission mechanism at radio wavelengths. 

\item \underline{Magnetic fields in the Local Universe:}
B-fields in external galaxies have only recently been measured using far-infrared (FIR) dust polarization. A legacy survey of nearby galaxies using the High-resolution Airborne Wideband Camera (HAWC+), installed on SOFIA \citep{Lopez-Rodriguez2020, Lopez-Rodriguez2022}, has traced the magnetic field structure in the plane of the sky in the cold gas for a small sample of nearby galaxies, on typical scales of $\sim300$~pc, not yet resolving the turbulence coherence scale.
The unparalleled spatial resolution and sensitivity of ALMA in full polarization has been able to observe on tens of pc-scales, capturing B-fields over the turbulence coherence length \citep[50--100~pc: ][]{Haverkorn2008} in the closest ($<10$~Mpc) galaxies.
Dust polarization (at 350~GHz) has been measured with ALMA
 in the inner 200~pc of the iconic starburst galaxy NGC~253 \citep{Belfiori2025}, 
 showing a local minima in the fractional polarization aligned along the galaxy midplane, where numerous superstar clusters --- massive and compact star-forming clusters --- are located.
These results highlight the interplay between magnetic fields, star formation, and outflows
in shaping the central region of NGC~253.
For the closest Ultra-Luminous Infrared galaxy, Arp~220, dust polarization has been observed showing a spiral-like magnetic field on the disc,  and the first detection of   
CO(3-2) G-K effect in an outflow has been reported 
\citep{Lopez-Rodriguez2026}.

Nevertheless, Arp~220 and NGC~253 are both nearby and unusually luminous galaxies, and it remains uncertain whether they are representative of most of the galaxy population in the Local Universe. Observations of a large sample of nearby galaxies sampling a variety of star formation properties and morphologies is needed to draw a full picture of the role of magnetic field in star formation processes and outflows in galaxies.  

\item \underline{Magnetic fields across cosmic times - Extragalactic studies:} In addition to detailed observations in the nearby Universe, polarized dust mapping has recently been successfully applied to early star-forming galaxies at redshifts up to $z\sim6$ (Geach et al. 2023, Chen et al. 2024; see Fig.~2).
Both studies reported polarization fractions close to the level of 1\%, similar to starburst galaxies in the nearby Universe. The resolved ordered magnetic fields extend up to $\sim$5~kpc, indicating that an efficient large-scale dynamo was already active in these early galaxies.
These studies heavily rely on local-Universe work to guide the interpretation of their results. Similarly, high-resolution imaging combined with a statistical sample is also key to advancing our understanding of the origin of B-fields and their role in global galaxy evolution. 

\end{itemize}

\textcolor{purple}{
{\bf 
Despite the significant progress we have made in our understanding of magnetic fields thanks to ALMA, its present sensitivity and polarization limited capabilities do not allow proper characterization of the magnetic field properties
} 
(in particular, the magnetic field strength estimates), sometimes yielding contradictory results, and generating vivid discussions on the feasibility of the statistical methods used \citep[e.g.][]{Liu2022}. In brief, it is clear that a major step forward is needed in the mm-wave spectro-polarimetry astronomy.
}

\section{Science Case for the 2040's}
\begin{tcolorbox}[boxsep=3pt,left=0pt,right=0pt,top=1pt,bottom=1pt,colframe=black,colback=lightcyan]
Understanding the magnetic fields with a new 3-dimensional perspective throughout the Local Universe and into the high-redshift universe, requires a  new generation (sub)-millimetre interferometer (hereafter ALMA2040), which can provide full spectro-polarimetry observations through all the frequency bands covered by ALMA, {\bf with  a high instrumental polarization purity}. 
Combined with enhanced spectral line and continuum sensitivity together with high spatial resolution, ALMA2040  will open an unprecedented view of the
origin and evolution of Galactic magnetic fields and the subsequent 
role of magnetic fields in molecular ISM dynamics, as well as star and planet formation, and during the later stages of  stellar evolution. 
\end{tcolorbox}
\noindent The specific goals for ALMA2040 should include:  
\begin{itemize}
    \item For the first time, \textbf{to obtain the 3-D topology and the strength of the B-field in different Galactic environments} (e.g., molecular envelopes around young stellar objects and  post-AGB stars, planet-forming disks, molecular outflows, PDRs) and in various types of galaxies, including the high-redshift universe. This scheme requires a coordinated approach that combines different polarization signatures: G-K and Zeeman effects, and dust polarization in combination with radiative transfer analysis. The present ALMA capabilities are too limited on all three to make such a meaningful analysis for more than a few, special objects.
    \item \textbf{To map out the evolution of the strength and structure of the magnetic field} across a large range of scales, from tens of pc down to few au, or in other words, \textbf{across the critical volume density regime for star formation, $10^4$--$10^9$~cm$^{-3}$,} in individual regions and across regions with sufficient precision and statistics to test star formation theories \citep{McKee2007, Girichidis2020, VazquezSemadeni2026}.
    \item \textbf{To test the driving mechanisms of molecular outflows and shock physics}. Elucidate how magnetic fields shape the evolution of evolved stars in the post-AGB to the Planetary Nebulae phase: 
    in particular, how do Alfv{\'e}n waves and coronal mass ejections contribute to the wind-driving mechanisms in the shocked extended atmosphere of evolved stars.
    How do magnetic fields regulate the launching jets during star formation and at the end of stellar evolution?
    \item 
    \textbf{Constraining the origin and developing timescale of  magnetic fields in galaxies by observing these fields across cosmic time}. To understand the role of the magnetic fields in the dynamics in extreme environments such as starburst galaxies, mergers, and AGN, but also in normal spiral galaxies and dwarf galaxies. A more precise measurement of the magnetic field strength will allow to much better constrain the magnetic field to gas density relationship in the cold, dense interstellar medium. This is of fundamental importance to test different MHD models of galaxy evolution \citep[e.g.,][]{Whitworth2025}.
\end{itemize}

We note that there are other specific ALMA2040 chapters closely related to this broad topic, such as {\it Probing ordered magnetic fields in galaxies across cosmic time} by Chen et al., {\it Tracking the flow of interstellar gas across scales: from galaxies to star clusters} by Duarte-Cabral et al, and {\it Star formation across the Galaxy} by Sánchez-Monge et al. In addition, other polarization mechanisms that will benefit from ALMA2040 are described in the {\it Multi-wavelength Polarization as a Probe of Planet Formation} chapter by Stephens et al. and solar polarization observations are briefly mentioned in {\it The energetic atmosphere of the Sun} by  Wedemeyer et al.


\begin{figure} 
\setcounter{figure}{1}
\includegraphics[width=14cm]{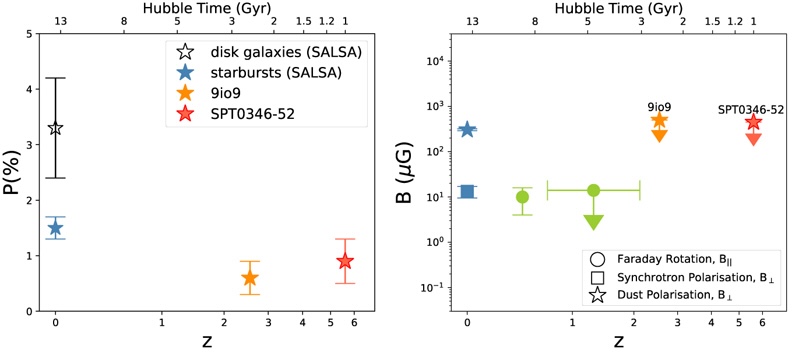}
\caption{\it 
Left: Measurements of dust polarization and magnetic fields at different redshifts. Measured polarization fraction as a function of redshift. Right: summary of the measured B-field strength in the distant galaxies, collected so far with different methodologies. Adapted from Figure~4 by \cite{Chen2024}.
{\bf ALMA2040 will allow to test the role of B-fields in galaxy evolution across cosmic times.}
\vspace{-0.4cm}
}
\end{figure}

\section{ALMA2040: Technical Requirements}\label{TechReq}

Present ALMA polarization capabilities are limited to band B1 up to B7, however, there are significant limitations on those bands.  The most severe limitation is for circular polarization, where a limit in the polarization fraction of 1.8\% or higher is only possible within the $1/10$ of the inner primary beam. In order to achieve the above science goals, ALMA2040 must provide: 
\begin{itemize}
    \item  \textbf{A 10x increase in instantaneous spectral line sensitivity in polarization [in Band 1–10 or 35–950~GHz]}, compared to current ALMA ($\sim$8x compared to anticipated ALMA Wideband Sensitivity Upgrade-- WSU). This increase is needed to achieve a spectral sensitivity to image key spectral lines with a SNR$\ge 1000$ for narrow lines ($<1$ km/s) with high fidelity images in all four Stokes parameter, so that detecting spectral line polarization (both Zeeman and G-K) becomes routine rather than an exception in full-polarization observations.
    \item  For observations of the complementary dust polarization, \textbf{a 10x increase in continuum sensitivity [in Band 1–10; 35–950GHz]}, compared to current ALMA ($\sim$5x compared to anticipated ALMA-WSU) is required. This will make it possible to track magnetic fields more comprehensively and consistently across all spatial scales probed by ALMA (in most observations, current sensitivity provides only an incomplete view of the B fields traced by dust polarization). The substantial sensitivity improvements anticipated for ALMA2040 will enable the detection of significantly fainter polarized emission, thereby directly mitigating sparse sampling—one of the largest sources of uncertainty inherent to the DCF method \citep[e.g., ][]{Liu2022}.
    \item {\bf Specific requirement for polarization: } (a) $<$0.1\% polarization purity (linear and circular) in continuum and spectral line modes; (b) Linear and circular polarization mosaics for both continuum and spectral lines observations. This will require a leakage correction across the full field of view; (c) Expand the polarization capabilities to all bands (e.g., to include \ci\ in B8 and B10, H$_2$O masers in B9, and high redshift \oii\ and \cii\ lines in bands B8, B9 and B10). Such extensive coverage across the wavelength range is essential for probing the polarized spectral energy distribution of dust emission, which is crucial for testing the composition of dust grains \citep{Hensley2023} as well as disentangling magnetic alignment of grains from other alignment and polarization mechanisms. 
\end{itemize}
These requirements can be achieved primarily by increasing the collecting area compared to ALMA today and disruptive developments in the optical design and receiver technology that significantly suppress the instrumental polarization, while at the same time providing exquisite image fidelity (through exquisite uv-coverage) across a range of angular scales. Innovative developments in receiver technology such as focal plane arrays, simultaneous multi-band capabilities, ultra-wide bandwidths (full band), will allow for mapping the strength and structure of magnetics fields from local planet-forming disks to high redshift galaxies.

\section{Synergies with current and future facilities}

The most straightforward synergies are with the two most ambitious next-generation facilities operating at longer wavelengths, the SKA and the ngVLA. Their broad multiwavelength coverage from radio through submillimeter will enable magnetic fields to be traced across every phase of the interstellar medium in external galaxies: ALMA2040 will constrain the magnetic field in the molecular component (via dust, Zeeman and G-K polarization), while the ngVLA and SKA will deliver comparable constraints in the atomic phase (using the Zeeman effect in the HI line) and in the ionized phase (through synchrotron emission and Faraday rotation).

Additional ALMA2040 synergies for magnetic-field studies involve current (sub)millimetre single-dish telescopes with full polarization capability—such as the IRAM 30m and the LMT 50m—as well as the future AtLAST and the far-IR PRIMA telescope. Together with ALMA2040, these facilities will make it possible to probe the magnetic field across a wide range of spatial scales, from molecular clouds down to disk scales.

\section*{Acknowledgements}
This work was developed as part of the broader ALMA2040 community effort. We acknowledge the Lorentz Center and NOVA for supporting the community workshops and coordination activities that helped initiate and shape the ALMA2040 White Paper process. We thank the anonymous reviewers for their constructive reports. \\

\medskip

\section*{Affiliations}

\vspace{0.2em}
{
\footnotesize{
\noindent $^1$ Institut de Ci\`encies de l'Espai (ICE-CSIC), Catalonia, Spain; 
$^2$ Jodrell Bank Centre for Astrophysics, The University of Manchester, UK;  
$^3$ Department of Physics and Astronomy, Chalmers University of Technology, SE-412 96 Gothenburg, Sweden;  
$^4$ Max-Planck-Institut für extraterrestrische Physik, Germany;  
$^5$ INAF – Osservatorio Astronomico di Arcetri, Italy;  
$^6$ Max-Planck-Institut f\"ur Astronomie (MPIA), Germany;  
$^7$ Leiden Observatory, Neetherlands;  
$^8$ NRAO, USA;
$^9$ JAO - ALMA, Chile;
$^{10}$ Observatoire de  Paris, France;  
$^{11}$ KASI, South Korea};
$^{12}$ Seoul National University, Seoul, South Korea;  
$^{13}$ National Tsing Hua University, Taiwan;  
$^{14}$ National Sun Yat-sen University, Taiwan; 
$^{15}$ Nanjing University, China;  
$^{16}$ University of South Carolina, USA;
$^{17}$ IPAG - Université Grenoble Alpes, France;  
$^{18}$ INAF – Istituto di Radioastronomia, Italy;  
$^{19}$IRyA, UNAM, México; 
$^{20}$ University College London, UK;  
$^{21}$ ESO, Germany;  
$^{22}$Nanjing University, China; 
$^{23}$CfA, Harvard \& Smithsonian, USA;
$^{24}$ CAB, INTA-CSIC, Spain;  
$^{25}$Queen's University, Canada;
$^{26}$Jet Propulsion Laboratory, California Institute of Technology, Pasadena, CA 91109, USA;
$^{27}$University of Tokyo, Japan;
$^{28}$ Institute of Science Tokyo, Japan
}

\vspace{0.7cm}


\bibliographystyle{apj_mod}
{\bibliography{references}}
\end{document}